\documentclass[aps,showpacs,prd,twocolumn]{revtex4}%
\usepackage{graphicx}
\usepackage{amssymb}
\usepackage{amsmath}
\usepackage{color}
\usepackage{float}
\usepackage{ulem}
\usepackage{accents}
\usepackage{graphicx}
\usepackage{graphicx}
\usepackage{amsfonts}
\usepackage{slashed}
\usepackage[colorlinks=true, pdfstartview=FitV,
linkcolor=blue,citecolor=blue,urlcolor=blue, breaklinks=true]{hyperref}%
\providecommand{\U}[1]{\protect\rule{.1in}{.1in}}
\begin{document}
\title{A dimension five Lorentz-violating nonminimal coupling for mesons in the KLZ model}
\author{V. E. Mouchrek-Santos}
\email{victor\_mouchrek@hotmail.com }
\author{Manoel M. Ferreira Jr}
\email{manojr.ufma@gmail.com}
\affiliation{Departamento de F\'{\i}sica, Universidade Federal do Maranh\~{a}o, Campus
Universit\'{a}rio do Bacanga, S\~{a}o Lu\'{\i}s - MA, 65080-805 - Brazil}
\author{Carlisson Miller}
\email{miller@ift.unesp.br}
\affiliation{Instituto de F\'{\i}sica Te\'orica, Universidade Estadual Paulista, Rua Dr.
Bento Teobaldo Ferraz, 271 - Bloco II, 01140-070 S\~ao Paulo, SP, Brazil}

\begin{abstract}
We investigate a dimension five Lorentz-violating (LV) nonminimal coupling between the neutral vector meson field strength and the
pion current in the context of the  Kroll-Lee-Zumino (KLZ) model. In this modified model we study the vector meson dominance (VMD) through of the imposing of new universality conditions between the couplings that comes from LV contributions. We also show that the LV nonminimal coupling leads to a linear momentum dependence in the timelike pion form factor, that is used to evaluate LV corrections for the muon anomalous magnetic moment. The experimental imprecision of such quantity allows us to obtain upper bound $\bar{g}\xi_{0} < 1.0 \mathbf{~}\mathrm{{GeV}}^{-1}.$ In addition, we also use the LV nonminimal coupling in the KLZ model to recalculate the $\rho$ decay rate. In this case the experimental imprecision is used to constrain the magnitude of the LV parameter at the level of $\bar{g}\xi_{0} < 1.9\times10^{-2}\mathbf{~}\mathrm{{GeV}}^{-1}$.

\end{abstract}

\maketitle

\section{Introduction}

The interaction of photon with the hadronic matter is one of the main aspects
of the hadronic physics and it is an important tool to learn about
QCD~\cite{Weise:1974bk}. Some applications involve the hadronic contribution
to the gyromagnetic ratio of the muon~\cite{Jegerlehner:2009ry}, the study of
the intrinsic structure of the nucleon~\cite{Kendall:1991np}, in-medium
modifications of hadrons~\cite{Leupold:2009kz}. At energies below 1 GeV, the
neutral vector mesons play an important role in this interaction,
being well described by the vector meson dominance model (VMD).
Such a model proposes that all interactions between photons and
hadrons, in this energy range, are mediated by the neutral vector mesons.

The VMD model was first studied by Sakurai~\cite{Sakurai,Sakurai:1960ju},
introducing the hadron-vector meson interactions analogously to the
electron-photon interactions in Quantum Electrodynamics. In
Ref.~\cite{Kroll:1967it} Kroll, Lee and Zumino proposed a renormalizable
Abelian quantum field theory for pions and neutral vector mesons. Despite the mass term for the gauge boson, the renormalizability is ensured
due to the coupling with a conserved
current~\cite{vanHees:2003dk,Ruegg:2003ps}. The great advantage of this theory is that it provides a quantum field theory justification for the
VMD model. Another important phenomenological study of the theory was the calculation of
the rho-meson self energy at the one-loop level, by Gale and
Kapusta~\cite{Gale:1990pn}. When one inserts this result in the VMD
expression for the electromagnetic pion form factor, one finds good agreement with experimental data in the timelike region. In addition, this form factor is consistent with the known Gounaris-Sakurai formula near to the rho-meson peak~\cite{Gounaris:1968mw,Davier:2005xq}.

The pion form factor is also directly related with the hadronic vacuum polarization contribution to
the muon anomalous magnetic moment~\cite{deTroconiz:2004yzs}. The muon magnetic moment has been experimentally measured with great precision~\cite{Brown:2001mga}: $10^{11} \times a_{\mu}({\rm{Exp.}}) = 116 592 080 \pm 60$. The great experimental precision also reveals a discrepancy with the theoretical value from the standard model~(SM)
at level of $2.3\sigma$ to $3.3\sigma$. Besides, the SM prediction to muon magnetic moment is conveniently separated into three parts: the electromagnetic (EM) part, weak (EW) part and the hadronic part (had). The electromagnetic and weak contributions have been determined with great accuracy~\cite{{Kinoshita:2004wi},{Jackiw:1972jz}}: $10^{11}\times a_{\mu
}(\mathrm{{QED}})=116584719\pm1.8,$ $10^{11}\times a_{\mu}(\mathrm{{EW}%
})=152\pm3.$ Combining these results, we find the {\textit experimental} value of the hadronic vacuum polarization contribution to $a_{\mu}$
\begin{eqnarray}
10^{11} \times a_{\mu}({\rm{had}}) = 7 209 \pm 60 .
\label{amuhad}
\end{eqnarray}

In the timelike region, the KLZ model was successful to describe the pion form
factor at tree level and giving the pion mean-squared radius $\langle r_{\pi
}^{2}\rangle|_{VMD}=0.39~\mathrm{{fm}^{2}}$, which is close to experimental
value~$\langle r_{\pi}^{2}\rangle|_{Exp}=0.439\pm0.008~\mathrm{{fm}^{2}}%
$~\cite{Patrignani:2016xqp}. The one-loop vertex
corrections to the tree level pion form factor were carried out in
Ref.~\cite{Dominguez:2007dm}.

Lorentz symmetry violation has been investigated with two main
purposes: as a possibility for the physics beyond the standard model and as a
precision programme that states to what extent Lorentz symmetry holds in
several sectors of physical interactions. One of the main frameworks for
addressing Lorentz violation is the gauge invariant and power-counting
renormalizable standard model extension (SME), developed by Colladay \&
Kostelecky~\cite{Colladay}. In the SME, Lorentz-violating (LV)
terms are fixed tensor-valued background fields, originated as vacuum
expectation values, that are coupled to the physical fields. Inside
the SME, LV effects were investigated in fermion systems
\cite{Fermion1,Fermion2}, in the CPT-odd electromagnetic sector
\cite{Adam1}, in the CPT-even electromagnetic sector
\cite{KM,Escobar}, in fermion-photon interactions
\cite{KFint,Interac,Vertex,QED}. Recently, Lorentz violation aspects were also
considered in the quark and hadrons interactions in the quiral regime
\cite{Quiral}, \cite{Noordmans}, \cite{Noordmans2}. Nonminimal extensions of the SME were developed, involving
higher derivatives and dimensional terms, both in the photon \cite{Kostelec1}
and fermion sector \cite{Kostelec2}. LV theories with
higher-dimensional terms were proposed inside other frameworks~\cite{Myers1,Marat}%
.

Nonminimal higher-dimensional couplings between fermions and photons
and not involving higher derivatives were also investigated in a CPT-odd
version \cite{NModd1}, addressed in several respects~\cite{NMmaluf,Radio,NMbakke,Pospelov}. Dimension five CPT-even
proposals were proposed in the context of the Dirac equation as well \cite{Jonas1}, with MDM and EDM effects being used to state
upper bounds at the level of $1$ part in $10^{20}~{\rm{(eV)}}^{-1}$ and $10^{24}~{\rm{(eV)}}^{-1}$,
respectively. A systematic investigation on LV NMCs of dimension five and six  was performed in Ref. \cite{Pospelov2}, \cite{Koste2}. Nonminimal couplings in the
electroweak sector \cite{Victor} and in a LV version of a scalar electrodynamics \cite{Josberg} were also proposed and constrained. Recently, new possibilities of dimension-6, dimension-7 and dimension-8 nonminimal couplings have been examined \cite{Koste3}, \cite{Jonas4}. For hadronic systems, in the VMD regime, the interaction between  photons and hadrons is played by vector mesons. In this scenario, it becomes senseful to
propose NMC between mesons as an extension of the NMC involving photons and
fermions of the QED, as a first step to examine NMC in QCD systems. This is the  main purpose of this paper.

The paper is organised as follows: In section~\ref{sec:LVVMD}, we revisit some aspects of the
KLZ model, discussing the VDM concept. We introduce an extra LV nonminimal coupling between the
neutral vector meson and the pion current, showing that it supports the VDM and the validity of the second VDM representation
in the universality limit. In section~\ref{sec:LVDECAY}, we study
the $\rho$-decay in the LV KLZ model, using experimental data to constrain the LV parameter. In
section~\ref{sec:LVPIONFF}, we evaluate the pion form factor corrected by the Lorentz
violation, examining the repercussions on the pion mean-squared radius and on the muon anomalous magnetic moment. In
section~\ref{sec:CONCLUSION} we present the conclusions and final remarks.

\section{Vector Meson Dominance in the KLZ model with LV nonminimal couplings }
\label{sec:LVVMD}

The KLZ lagrangian, involving the pions and vector mesons, is originally given by%
\begin{align}
\mathcal{L}  & =-\frac{1}{4}G^{\mu\nu}G_{\mu\nu}+\frac{1}{2}m_{\rho}^{2}%
\rho^{\mu}\rho_{\mu}+\partial_{\mu}\pi^{\ast}\partial^{\mu}\pi-m_{\pi}^{2}%
\pi^{\ast}\pi\nonumber\\
& +g_{\rho\pi\pi}\rho_{\mu}J_{\pi}^{\mu}
,\label{KLZLIV}%
\end{align}
with $m_{\pi}$ and $m_{\rho}$ being the pion and $\rho$-meson masses,
respectively. The complex field $\pi^{\ast}$ represents the charged
pions, $\pi^{\pm}$, the $\rho^{\mu}$ field describes the neutral vector
meson, $\rho^{0}$, while $J_{\pi}^{\mu}=i(\pi\partial_{\mu}\pi^{\ast}-\pi^{\ast}\partial_{\mu}\pi)$ is the the pion current.

The authors realized the vector meson dominance in the model via a mixing between the photon and the $\rho$ meson through a term of the form $F_{\mu\nu}G^{\mu\nu}$~\cite{Kroll:1967it}. This way, omitting the pion fields, the relevant terms are
\begin{align}
\mathcal{L}_{VMD} &  =-\frac{1}{4}F_{\mu\nu}F^{\mu\nu}-\frac{1}{4}G_{\mu\nu
}G^{\mu\nu}-\frac{1}{2}\frac{e}{g_{\rho}}F_{\mu\nu}G^{\mu\nu}\nonumber\\
&  +\frac{1}{2}m_{\rho}^{2}\rho_{\mu}\rho^{\mu}+g_{\rho\pi\pi}\rho_{\mu}%
J_{\pi}^{\mu}+eA_{\mu}J_{\pi}^{\mu},\label{KLZVMD1}%
\end{align}
where the electromagnetic field is directly coupled to the pion
current, $J_{\pi}^{\mu},$ but not to the neutral vector meson, which
only interacts with the photon by the field strength coupling. The
most natural coupling, $\rho_{\mu}A^{\mu}$, cannot be considered
because it leads to an imaginary mass contribution for the photon when one
calculates the dressed photon propagator~\cite{Sakurai:1960ju}. From Lagrangian~(\ref{KLZVMD1}), we derive the equations of motion for the $\rho $-meson and photon fields. After decoupling these equations, and keeping only terms up to linear order in the coupling constant $e$, the
electromagnetic current is written as
\begin{equation}
J_{em}^{\mu}=\frac{e}{g_{\rho}}m_{\rho}^{2}\rho^{\mu}+e\Big(\frac{g_{\rho
\pi\pi}}{g_{\rho}}-1\Big)J_{\pi}^{\mu},\label{current1}%
\end{equation}
which receives a hadronic contribution and shows that the gauge invariance is related to the $\rho$ meson transversality. This model characterizes the first version of vector meson dominance and it has a elegant form because the electromagnetic gauge invariance is explicitly fulfilled.

\begin{figure}
  \centering
  \includegraphics[width=9.0cm]{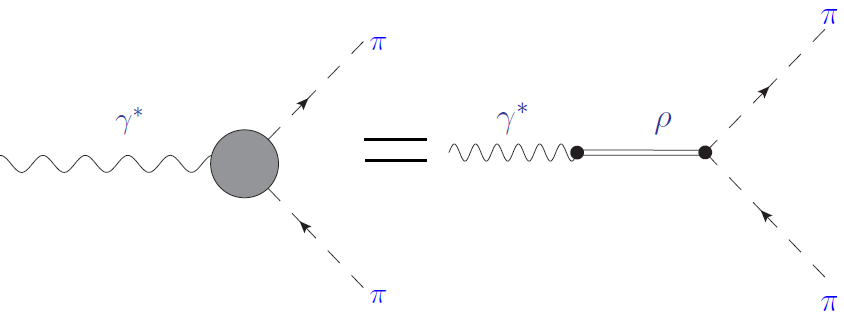}\\
  \caption{ Interaction between photon and pion mediated by a $\rho$ meson }
  \label{fig1}
\end{figure}

In the particular limit $g_{\rho\pi\pi}=g_{\rho}$, the photon converts entirely into a neutral $\rho$ meson, once that the pion current decouples from the electromagnetic current, existing no longer direct coupling between the photon and the pion current. This scenario becomes clearer by replacing the following transformations
\begin{equation}
A_{\mu}=A_{\mu}^{\prime}/\kappa,\text{ \ }\rho_{\mu}=\rho_{\mu}^{\prime}%
-\frac{e}{g_{\rho}}A_{\mu},\text{ \ }e=e^{\prime}/\kappa,\label{KLZTF}%
\end{equation}
in the lagrangian~(\ref{KLZVMD1}), with $\kappa=\sqrt{1-\left(  e/g_{\rho }\right)  ^{2}},$ which provides
\begin{align}
\mathcal{L}_{VMD2} &  =-\frac{1}{4}F^{\prime\mu\nu}F_{\mu\nu}^{\prime}%
+m_{\rho}^{2}\frac{e^{2}}{2g_{\rho}^{2}}A_{\mu}^{\prime}A^{\prime\mu}-\frac
{1}{4}G^{\prime\mu\nu}G_{\mu\nu}^{\prime}\nonumber\\
&  +\frac{1}{2}m_{\rho}^{2}\rho_{\mu}^{\prime}\rho^{\prime\mu}-m_{\rho}%
^{2}\frac{e}{g_{\rho}}\rho_{\mu}^{\prime}A^{\prime\mu}+g_{\rho\pi\pi}\rho
_{\mu}^{\prime}J_{\pi}^{\mu}.\label{KLZVMD2}%
\end{align}
Notice that the direct interaction term between the photon and the pion current,
$e\left[  1-g_{\rho\pi\pi}/g_{\rho}\right]  A_{\mu}J_{\pi}^{\mu},$ has
vanished in the limit $g_{\rho\pi\pi}=g_{\rho}$, which now is mediated only by the
neutral $\rho$-meson. Now the electromagnetic current is entirely read
in terms of the neutral $\rho$-meson,
\begin{equation}
J_{em}^{\prime\mu}=\frac{e}{g_{\rho}}m_{\rho}^{2}\rho^{\prime\mu}.
\label{current2}
\end{equation}
This limit is well known as a universality condition, which allows to obtain the interaction between the photon and pion only mediated by a vector meson, as illustrated in Fig.~\ref{fig1}. The Lagrangian~(\ref{KLZVMD2}) reads as the second
version of the vector meson dominance and it is more used because the propagators are diagonals. However, the electromagnetic gauge invariance is not explicit anymore, due to the photon mass and the mixing terms.

We should now introduce Lorentz violation in this kind of model
without spoiling the VMD, however. To do that, we add in
Lagrangian~(\ref{KLZVMD1}) two dimension five nonminimal couplings,%
\begin{equation}
\mathcal{L}_{LV}=\bar{g} J_{\pi}^{\mu}G_{\mu\nu}\xi^{\nu}+\tilde{g}J_{\pi}^{\mu
}F_{\mu\nu}\xi^{\nu},\label{LVNMC1}
\end{equation}
which represent unusual interactions of the field strengths with the
pion current and a fixed LV background,
$\xi^{\nu},$ stating a preferred direction in spacetime and violating Lorentz
symmetry. Such a coupling has mass dimension equal to
$-1,$ so that this theory becomes nonrenormalizable by
power-counting. According to Wilson renormalization group, it can be interpreted as an effective theory with a cutoff.

Considering the two terms in (\ref{LVNMC1}), there appears an additional
term in the electromagnetic current (\ref{current1}),
\begin{equation}
J_{em,LV}^{\mu}=\left[  \frac{e}{g_{\rho}}\bar{g}-\tilde{g}\right]
\xi^{\alpha}\partial_{\alpha}J_{\pi}^{\mu},\label{EMCurrentVMD1}%
\end{equation}
and new terms in Lagrangian (\ref{KLZVMD2}),
\begin{equation}
\mathcal{L}_{LV}^{\prime}=\bar{g}\xi^{\mu}G_{\mu\nu}^{\prime}J_{\pi}^{\mu
}+\left[\tilde{g} - \bar{g} \frac{e}{g_{\rho}}\right]  \xi^{\mu}F_{\mu\nu}J_{\pi
}^{\mu},\label{LLV2}%
\end{equation}
when the transformations~(\ref{KLZTF}) are implemented. This way, the Lorentz violation term brings an extra contribution for the EM current also associated to pion current. In order to keep the photon converting into a $\rho$ meson we impose a new universality condition for the couplings that comes from the LV terms, namely, $\tilde{g} g_{\rho}=e \bar{g}$. As a consequence, the nonminimal LV interaction between photon and pion current vanishes and only remains $\mathcal{L}_{LV}^{\prime}=\bar{g}\xi^{\mu}G_{\mu\nu}^{\prime}J_{\pi}^{\mu}$. This is second version of VMD with Lorentz violation and its full the Lagrangian to be regarded from now on is
\begin{align}
\mathcal{L}_{LVVMD2} & = -\frac{1}{4}F^{\prime\mu\nu}F_{\mu\nu}^{\prime}%
+m_{\rho}^{2}\frac{e^{2}}{2g_{\rho}^{2}}A_{\mu}^{\prime}A^{\prime\mu}-\frac
{1}{4}G^{\prime\mu\nu}G_{\mu\nu}^{\prime}\nonumber\\
&  +\frac{1}{2}m_{\rho}^{2}\rho_{\mu}^{\prime}\rho^{\prime\mu}-m_{\rho}%
^{2}\frac{e}{g_{\rho}}\rho_{\mu}^{\prime}A^{\prime\mu}+g_{\rho\pi\pi}\rho
_{\mu}^{\prime}J_{\pi}^{\mu}\nonumber\\
&  + g_{1} \xi^{\mu}G_{\mu\nu}^{\prime}J_{\pi}^{\mu},\label{KLZVMD3}%
\end{align}
whose EM current is given in~(\ref{current2}). Notice that we have not included the second order contribution involving the usual coupling, $g_{\rho\pi\pi}^{2}\rho_{\mu}\rho^{\mu}\pi^{\ast}\pi$, since it is not relevant in the demonstration of the vector meson dominance. But, it will be included to address the vacuum polarization contribution for the pion form factor. There is also a second order term in the LV coupling, $\bar{g}^{2}\xi^{\mu}\xi_{\nu} \rho_{\mu\alpha}\rho^{\nu\alpha}\pi^{\ast}\pi$, which will be neglected. Its worth to mention that, in the absence of the LV term, despite
the nonvanishing mass term, the $\rho$-meson field still obeys the
transversality condition, $\partial_{\mu}\rho^{\mu}=0$. This is possible
because the neutral vector meson is coupled to a conserved
current~\cite{Kroll:1967it,Lowenstein:1972pr}.

\section{\ LV corrections to the decay rate of the decay $\left(  \rho
^{0}\rightarrow\pi^{-}\pi^{+}\right)  $}
\label{sec:LVDECAY}

In this section, we examine just the effect of the nonminimal coupling, $\bar
{g}J_{\pi}^{\mu}G_{\mu\nu}\xi^{\nu}$, on the decay of the neutral $\rho$-meson
into two pions. The photons play no role in the process, hence, the lagrangian term associated to it is written as
\begin{equation}
\mathcal{L}_{\rho\pi\pi}=\left[  g_{\rho\pi\pi}\rho_{\mu}\left(  x\right)
+\bar{g}G_{\mu\nu}\xi^{\nu}\right]  J_{\pi}^{\mu},
\label{LRPP}%
\end{equation}
and the corresponding scattering matrix is given by
\begin{align}
S &  =-i\int d^{4}x\left[  g_{\rho\pi\pi}\rho_{\mu}\left(  x\right)  +\bar
{g}G_{\mu\nu}\xi^{\nu}\right]  J_{\pi}^{\mu}\nonumber\\
&  =S_{0}+S_{LV\left(  1\right)  }+S_{LV\left(  2\right)  }.\label{S}%
\end{align}
We then propose plane wave expansions, $\rho_{\mu}\left( x\right)
=\left(  2V q_{0}\right)  ^{-1/2}\varepsilon_{\mu}\left( q,\lambda\right)
\exp\left(  -iq\cdot x\right)  $, $\pi=\left(  2V p_{0}\right)^{-1/2}%
\exp\left(  -ip\cdot x\right)  $, $\pi^{\ast}=\left(  2Vp_{0}^{\prime}\right)
^{-1/2}\exp\left(  ip^{\prime}\cdot x\right)  $, where $q,p,p^{\prime}$ stand
for the $\rho^{0}$ vector meson and the emerging
pion/antipion four-momenta, respectively. The zero order and first order
contributions in the LV parameters are
\begin{align}
S_{0} &  =g_{\rho\pi\pi}\left(  2\pi\right)  ^{4}\frac{\delta^{4}\left(
p^{\prime}-p-q\right)  }{\left[  8V^{3}q_{0}p_{0}p_{0}^{\prime}\right]
^{1/2}}M_{0},\label{S0}\\
S_{LV\left(  a\right)  } &  =\bar{g}\left(  2\pi\right)  ^{4}\frac{\delta
^{4}\left( p^{\prime}-p-q\right)  }{\left[  8V^{3}q_{0}p_{0}p_{0}^{\prime
}\right]  ^{1/2}}M_{LV\left(  a\right)  },
\end{align}
with the usual amplitude,
\begin{equation}
\left.  M_{0}=\varepsilon_{\mu}\left( q,\lambda\right)  \left( p^{\prime\mu
}-p^{\mu}\right)  ,\right.
\end{equation}
and $a=1,2$ representing the two LV contributions, which involve
\begin{align}
&  \left.  M_{LV\left(  1\right)  }=-\xi^{\nu}q_{\mu}\varepsilon_{\nu}\left(
q,\lambda\right)  \left(  p^{\prime\mu}-p^{\mu}\right)  ,\right. \nonumber  \\
&  \left.  M_{LV\left(  2\right)  }=\xi^{\nu}q_{\nu}M_{0}.\right.
\label{MLV1b}%
\end{align}
The decay rate for the tree-level process $\left( \rho
^{0}\rightarrow\pi^{-}\pi^{+}\right)  $ is usually given as
\begin{equation}
\Gamma=\frac{1}{T}V\int\frac{d^{3}q}{\left(  2\pi\right)  ^{3}}V\int%
\frac{d^{3}q^{\prime}}{\left(  2\pi\right)  ^{3}}\frac{1}{3}\sum_{\lambda
}\left\vert S\right\vert ^{2},\label{gama}%
\end{equation}
with $S$ defined in Eq. (\ref{S}). The squared matrix amplitude is
\begin{equation}
\left\vert S\right\vert ^{2}=S_{0}S_{0}^{\dagger}+S_{0}S_{LV\left(  1\right)
}^{\dagger}+S_{LV\left(  1\right)  }S_{0}^{\dagger}+S_{0}S_{LV\left(
2\right)  }^{\dagger}+S_{LV\left(  2\right)  }S_{0}^{\dagger},\label{S2}%
\end{equation}
in first order in the LV parameters.\textbf{ }Substituting Eq. (\ref{S2}) in
Eq. (\ref{gama}), ones achieves
\begin{align}
\Gamma_{ll} &= \Gamma_{S_{0}S_{0}^{\dagger}}+\Gamma_{S_{0}S_{LV\left(
1\right)  }^{\dagger}}+\Gamma_{S_{LV\left(  1\right)  }S_{0}^{\dagger}}
+\Gamma_{S_{0}S_{LV\left(  2\right)  }^{\dagger}} \nonumber \\
&+\Gamma_{S_{LV\left( 2\right) } S_{0}^{\dagger}}. \label{VSL1.19.2}
\end{align}
The first term, $\Gamma_{S_{0}S_{0}^{\dagger}},$ is the decay rate for the
Lorentz invariant usual process $\left(  \rho\rightarrow\pi^{-}\pi^{+}\right)
$, that is,
\begin{equation}
\Gamma_{S_{0}S_{0}^{\dagger}}=\frac{g_{\rho\pi\pi}^{2}}{48\pi}\frac{\left(
m_{\rho}^{2}-4m_{\pi}^{2}\right)  ^{3/2}}{m_{\rho}^{2}}\Theta\left(  m_{\rho
}-2m_{\pi}\right)  .
\end{equation}
As a consequence of the current conservation, due to the presence of the
momentum $q_{\mu}$ in Eq. (\ref{MLV1b}), one has $\Gamma
_{S_{0}S_{LV\left(  1\right)  }^{\dagger}}=0,$ $\Gamma_{S_{LV\left(  1\right)
}S_{0}^{\dagger}}=0.$ The nonnull first order LV contribution stems
from the pieces $\Gamma_{S_{0}S_{LV\left(  2\right)}^{\dagger}}$and
$\Gamma_{S_{LV\left(  2\right)  }S_{0}^{\dagger}}$, which can be read as
\begin{align}
\Gamma_{S_{0}S_{LV\left(  2\right)  }^{\dagger}} &  =\Gamma_{S_{LV\left(
2\right)  }S_{0}^{\dagger}}=\left(  \xi\cdot q\right)  \frac{\bar{g}g_{\rho
\pi\pi}}{48\pi k_{0}}\frac{\left(  m_{\rho}^{2}-4m_{\pi}^{2}\right)  ^{3/2}%
}{m_{\rho}}\nonumber\\
&  \times\Theta\left(  m_{\rho}-2m_{\pi}\right)  .\label{decayR}%
\end{align}
where $\xi\cdot q=\xi_{0}q_{0}-\boldsymbol{\xi}\cdot\mathbf{q.}$ Therefore, the total decay rate, $\Gamma=\Gamma_{S_{0}S_{0}^{\dagger}}%
+\Gamma_{S_{0}S_{LV\left(  2\right)  }^{\dagger}}+\Gamma_{S_{LV\left(
2\right)  }S_{0}^{\dagger}},$ is now written as
\begin{align}
\Gamma &  =\frac{g_{\rho\pi\pi}^{2}}{48\pi}\frac{\left(  m_{\rho}^{2}-4m_{\pi
}^{2}\right)  ^{3/2}}{m_{\rho}^{2}}\left[  1+\frac{2m_{\rho}}{g_{\rho\pi\pi}%
}\left(  \bar{g}\xi_{0}\right)  \right]  \nonumber\\
&  \times\Theta\left(  m_{\rho}-2m_{\pi}\right)  ,\label{decayR2}%
\end{align}
where we have used $q_{0}^{2}=m_{\rho}^{2}$ and $\left( \xi\cdot q\right)
=\xi_{0}m_{\rho}$, since we are adopting the point of view of the rest frame
of the $\rho^{0}$ vector meson.

In accordance with Ref.~\cite{Patrignani:2016xqp}, the meson $\rho$ decay rate is
$\Gamma=\left(  149.1\pm0.8\right)~\mathrm{{MeV}}$ or $\Gamma=149.1\left(
1\pm0.005\right)~\mathrm{{MeV}}$ and the $\rho$ mass is $m_{\rho}=775.50~\mathrm{{MeV}%
}$~\cite{Patrignani:2016xqp}. The coupling is $g_{\rho\pi\pi}$ can be estimated by the universality condition and it is given by $g_{\rho\pi\pi}=6.0$~\cite{Djukanovic:2004mm}. So, we impose that the LV contribution can not be larger than the experimental imprecision of the measurement. At this way,
it holds $2m_{\rho}\left(  \bar{g}\xi_{0}\right)  /g_{\rho\pi\pi}%
<5.0\times10^{-3}$, which leads to the upper bound
\begin{align}
\left(  \bar{g}\xi_{0}\right)   &  <1.9\times10^{-2}~\left(  \mathrm{{GeV}%
}\right)  \mathrm{^{-1}.}%
\label{bound1}
\end{align}

In the framework of the SME, the LV tensor backgrounds are considered
fixed in spacetime, that is, in the Sun's frame. In the Earth frame~\cite{Jonas1,Sideral},~these coefficients undergo sidereal variations. Thus, it is suitable to translate the bounds from the
Earth-located Lab's RF at the colatitude $\chi$, rotating around the Earth's axis with angular velocity $\Omega$, to the Sun\'{}s frame. For experiments up to a few weeks long, the transformation law for a rank-1 tensor, $A_{\mu}$, is merely a spatial rotation, $A_{\mu}%
^{T}=R_{\mu\alpha}A_{\alpha},$ where the label $T$ indicates
the quantity measured in the Sun's frame, and $R_{0i}=R_{i0}=0$ and
$R_{00}=1.$ Thus, four vector time-components are not modified,%
$\ \ A_{0}^{T}=A_{0}$, so that the upper bound (\ref{bound1}) could be
equally written in the Sun's frame. However, the situation is not so simple,
as pointed out in Ref.~\cite{Pion} (for pion decays), once the decay rate of
the process $\rho^{0}\rightarrow\pi^{-}\pi^{+}$ is being considered
in the rest frame of the decaying meson, not in the Lab (Earth) frame, where
the measurements are performed. In order to take into account this point, one
option is to translate the upper bounds (\ref{bound1}), associated with an
evaluation at the vector boson rest frame, directly to the Sun\'{}s frame, with the boost
$\xi^{0}=\gamma_{z}(\xi_{T}^{0}+\alpha^{i}\xi_{T}^{i})$,
where $\gamma_{z}=\gamma(v_{z})$ is the Lorentz factor, $v_{z}$ is the meson velocity in the Sun\'{}s frame, $\alpha^{i}=v_{z}^{i}$\textbf{ }$/c$, approximately its
velocity in LAB frame. In this case, the upper bounds (\ref{bound1}) can be read in the Sun\'{}s frame as
\begin{equation}
\left\vert \bar{g}(\xi_{T}^{0}+\alpha^{i}\xi_{T}^{i})\right\vert
\lesssim1.9\times10^{-2}\mathbf{~}\left(  \mathrm{{GeV}}\right)
^{-1},\label{boundS1}%
\end{equation}
since $\gamma_{z}\simeq1$ in the decay experiments with available CM energy not much larger than $m_{\rho}=775.50~\mathrm{{MeV}%
}$.

\section{LV corrections to the pion form factor}
\label{sec:LVPIONFF}

As we have pointed out in the introduction, one of the most important aspects of the KLZ
model is its ability for providing the electromagnetic pion form factor. Experimentally, the timelike
pion form factor can be measured through the process $e^{+}e^{-}\rightarrow
\gamma\rightarrow\pi^{+}\pi^{-}$ and the spacelike pion form factor can be measured through the process $e^{-}\pi^{-}\rightarrow \gamma\rightarrow e^{-}\pi^{-}$~\cite{Dominguez:2007dm,Lomon:2016eyp}. In the timelike region, it holds
\begin{equation}
\langle\pi(p^{\prime})\pi(p)|J_{em}^{\mu}|0\rangle=(p^{\prime}-p )^{\mu}F_{\pi}.
\end{equation}

This matrix element can be evaluated directly from the KLZ model with Lorentz
violation. Following Lagrangian (\ref{KLZVMD3}), we obtain the following expression for
the pion form factor at tree level:
\begin{equation}
F_{\pi}(q)=-\frac{m_{\rho}^{2}}{q^{2}-m_{\rho}^{2}}(1+\frac{\bar{g}}%
{g_{\rho\pi\pi}}\xi^{\mu}q_{\mu}),\label{FFVMD2}%
\end{equation}
where we have omitted the prime index of the fields and the transfer momentum is $q^{\mu}=( p^{\prime} + p )^{\mu}$. Notice the appearance of a linear contribution in the
photon momentum, contracted with the background vector, $\xi^{\mu},$
constituting an anisotropy source for the pion form factor.

\begin{figure}
  \centering
  \includegraphics[width=9.0cm]{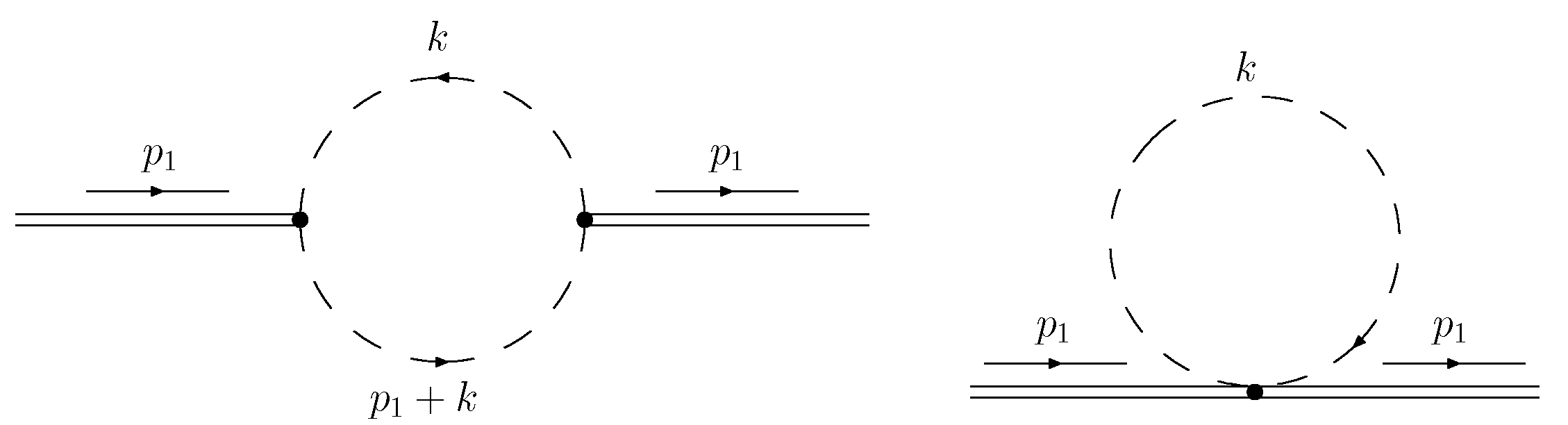}\\
  \caption{ Diagrams relevant for the $\rho$ self-energy }
  \label{fig3}
\end{figure}

As mentioned before, beyond the tree-level analysis, the form factor in Eq.~(\ref{FFVMD2}) gains a pole at the $\rho$ mass, stemming from the $\rho$-meson self energy evaluations. In the actual version of VMD, the LV new vertex also brings a correction for the $\rho$-meson self-energy. The relevant diagrams are shown in Fig.~\ref{fig3}, showing only the vacuum polarization contribution. This way, the $\rho$ propagator is corrected by the vacuum polarization contribution, so that pion form factor~(\ref{FFVMD2}) becomes
\begin{equation}
F_{\pi}(q)=-\frac{m_{\rho}^{2} + \Pi_{\rho}(0)}{q^2 - m_{\rho}^{2} - \Pi_{\rho}(q)}(1+\frac{\bar{g}}%
{g_{\rho\pi\pi}}\xi^{\mu}q_{\mu}).\label{FFVMD3}%
\end{equation}

The vacuum polarization receives contribution from the usual vertex and that from the LV correction, $\bar{g}G_{\mu\nu}\xi^{\nu}$. From Fig.~\ref{fig3}, one observes that this quantity is quadratic in the usual coupling. Thus, we have to include the second order contribution which comes from the term $g^{2}_{\rho\pi\pi}\rho_{\mu}\rho^{\mu}\pi^{\ast}\pi$. The presence of the LV coupling will bring a crossed contribution, like $g_{\rho\pi\pi}\bar{g}$, which is a first order LV correction. In addition, there is a second order LV correction which will be neglected.

Although the LV interaction term depends on the $\rho$-meson momentum, it does not contribute for the pion loop integration. Therefore, it is straightforward to find the following expression for the LV vacuum polarization contribution
\begin{equation}
\Pi_{\rho}(q) = \Big( 1 + \frac{2\bar{g}\xi^{\alpha}}{g_{\rho\pi\pi}} q_{\alpha} \Big) \Pi_{\rho}^{(0)}(q^2),
\label{LVVP1}
\end{equation}
where only first order LV terms are considered. The $\Pi_{\rho}^{(0)}(q^2)$ function is just the usual vacuum polarization contribution, which can be similarly achieved as done in a massive photon scalar electrodynamics~\cite{Dominguez:2007dm}, that is
\begin{eqnarray}
\Pi_{\rho}^{(0)}(q^2) = A q^2 + B - Gq^2D(q) \Big[ i \pi-LN \Big],
\label{LVVP2}
\end{eqnarray}
where $LN=\ln \Big\vert \frac{\sqrt{(1-4m_{\pi}^2/q^2)} + 1}{\sqrt{(1-4m_{\pi}^2/q^2)}-1}\Big\vert$, $G=\frac{1}{3} \frac{g_{\rho\pi\pi}^2}{(4\pi)^2}$ and $D(q)=\Big(1-{4m_{\pi}^2}/{q^2}\Big)^{3/2}$. This form is valid above the production threshold $q^2 > 4 m_{\pi}^2$. Below this threshold $0 < q^2 < 4 m_{\pi}^2$, we have
\begin{eqnarray}
\Pi_{\rho}^{(0)}(q^2) &=& - \frac{g_{\rho\pi\pi}^2}{24\pi^2} q^2 \Big[
\Big(\frac{4m_{\pi}^2}{q^2} - 1\Big)^{3/2} \arcsin(\frac{\sqrt{q^2}}{2m}) \nonumber \\
&&+ \frac{4}{3}
- \frac{4m_{\pi}^2}{q^2} \Big] + A q^2 + B,
\label{LVVP3}
\end{eqnarray}
where the constants $A$ and $B$ are determined by the vacuum polarization at $\rho$ mass and zero momentum, respectively, being given by
\begin{equation}
A=-G\Big[D(q)LN+8\frac{m_{\pi}^{2}}{m_{\rho}^{2}}\Big],\text{ }B=\Pi_{\rho
}^{(0)}(0)=8Gm_{\pi}^{2}.
\end{equation}

In order to keep LV effects only at first order, we expand the LV pion form factor~(\ref{FFVMD3}) using the LV vacuum polarization~(\ref{LVVP1}) and we have
\begin{equation}
F_{\pi}(q) = F_{\pi}^{(0)}(q^2) \Big(1 + C(q^2)\frac{\bar{g}}{g_{\rho\pi\pi}}\xi^{\mu}q_{\mu} \Big),
\label{FFVMD4}%
\end{equation}
where the pion form factor in the absence of LV, $F_{\pi}^{(0)}(s)$, and the factor $C(q^2)$, are
\begin{equation}
F_{\pi}^{(0)}(q^2)= -\frac{m_{\rho}^{2}+\Pi_{\rho}^{(0)}(q^2)}{q^2-m_{\rho}^{2}-\Pi_{\rho}^{(0)}(q^2)},
\label{FFVMD5}%
\end{equation}
\begin{equation}
C(q^2) = \frac{q^2 - m_{\rho}^{2} + \Pi_{\rho}^{(0)}(q^2)}{q^2 - m_{\rho}^{2} - \Pi_{\rho}^{(0)}(q^2)}.\label{C}%
\end{equation}
The consequences of the linear momentum dependence in~(\ref{FFVMD4}) can be seen by looking it at low momenta. This way, the LV pion form factor allows the expansion
\begin{equation}
F_{\pi} (q)= 1 + C(0)\frac{\bar{g}}{g_{\rho\pi\pi}}\xi^{\mu}q_{\mu}
+ \frac{q^2}{ m_{\rho}^{2} + \Pi_{\rho}^{(0)}(0)}\ldots .
\label{FFVMD6}%
\end{equation}

Notice that the well known pion mean-squared radius, the coefficient of $q^2$, is not affected by LV factor. On the other hand, the LV interaction brings a nonzero pion mean radius, the coefficient of the linear momentum. This might have a deep consequence in the properties of the pion, as playing the role of
a nonzero electric dipole momentum for the pion. We will leave this possibility to be discussed in a future work.

We work in the center mass frame where $p^{\prime\mu}=(E,+\vec{p})$ and $p^{\mu}=(E,-\vec{p})$. This way, we obtain the relation $q^2 = s = 4E^2$ and naturally only the time direction of the Lorentz vector $\xi^{\mu}$ takes place. So, the LV pion form factor~(\ref{FFVMD4}) reads
\begin{eqnarray}
F_{\pi}(s)= F_{\pi}^{(0)}(s) \Big( 1 + C(s) \frac{\bar{g}\xi_{0}}{g_{\rho\pi\pi}} \sqrt{s} \Big).
\label{FFVMD7}%
\end{eqnarray}
where LV effects are carried just by the time component of the LV background vector, $\xi_0$.

\subsection{LV correction to muon anomalous magnetic moment.}

Another possible route for constraining the magnitude of the LV nonminimal coupling is the hadronic contribution of the
$\pi^{+}\pi^{-}$ timelike channel to muon anomalous magnetic moment. At lowest order~\cite{deTroconiz:2004yzs}, the hadronic contribution to this quantity is
\begin{eqnarray}
a_{\mu}^{had} = \int_{4m_{\pi}^2}^{\infty} ds K(s) R(s),
\label{AMM}
\end{eqnarray}
where the Kernel function is given by
\begin{eqnarray}
K(s) = \frac{\alpha^2}{3\pi^2 s} \int_{0}^{1} dx \frac{x^2(1-x)}{x^2 + (1-x)s/m_{\mu}^{2}} ,
\end{eqnarray}
the fine structure constant is $\alpha=1/137$ and $m_{\mu}$ is the muon mass. The function $R(s)$ is the ratio of the cross sections for $e^{+}e^{-}$ annihilation into hadrons over annihilation into muons, at lowest order:
\begin{eqnarray}
R(s) = \frac{\sigma (e^{+}e^{-}\rightarrow {\rm{hadrons}})}{\sigma (e^{+}e^{-}\rightarrow \mu^{+}\mu^{-})}.
\label{Ratio}
\end{eqnarray}

The cross sections for $e^{+}e^{-}$ annihilation into muons is well known,
\begin{eqnarray}
\sigma (e^{+}e^{-}\rightarrow \mu^{+}\mu^{-}) = \frac{4\pi\alpha^2}{3s}.
\end{eqnarray}
At low energies $s\le s_{0} = 0.8~\rm{GeV}^2$, where just the light quarks are relevant, the cross sections for $e^{+}e^{-}$ annihilation into hadrons receives just the two-pion contribution and it can be expressed as
\begin{eqnarray}
\sigma (e^{+}e^{-}\rightarrow \pi^{+}\pi^{-}) = \frac{\pi\alpha^2}{3s} \Big(1 - \frac{4m_{\pi}^2}{s} \Big)^{3/2} \vert F_{\pi}(s) \vert^2.
\end{eqnarray}
This way, the hadronic contribution to muon anomalous magnetic moment~(\ref{AMM}) is written as
\begin{eqnarray}
a_{\mu}^{had,\pi} = \frac{1}{4}\int_{4m_{\pi}^2}^{s_{0}} K(s) \Big(1-\frac{4m_{\pi}^2}{s}\Big)^{3/2}
\vert F_{\pi}(s) \vert^{2} ds .\label{AMM1}
\end{eqnarray}

Now, it is convenient to rewrite the muon anomalous magnetic moment~(\ref{AMM1}) in one contribution without LV and another at first order in LV. Then, using the LV pion form factor~(\ref{FFVMD7}), we have
\begin{eqnarray}
a_{\mu}^{had,\pi} = a_{\mu}^{(0)} + \frac{2\bar{g}\xi_0}{g_{\rho\pi\pi}} a_{\mu}^{(1)} ,
\label{AMM2}
\end{eqnarray}
where $a_{\mu}^{(0)}$ is the hadronic contribution for the muon anomalous magnetic moment in the absence of LV, namely, using $F_{\pi}^{(0)}(s)$. The $a_{\mu}^{(1)}$ term, which receives LV contribution inside the parenthesis in Eq.~(\ref{FFVMD7}), is given by
\begin{eqnarray}
a_{\mu}^{(1)} = \frac{1}{4}\int_{4m_{\pi}^2}^{s_{0}} K(s) \Big(1-\frac{4m_{\pi}^2}{s}\Big)^{3/2}
{\rm{Re}}C(s) \sqrt{s} \vert F_{\pi}^{(0)}(s) \vert^{2} ds . \nonumber \\
\label{AMM3}
\end{eqnarray}

With the muon mass $m_{\mu}=0.1056$ GeV~\cite{Patrignani:2016xqp}, the numerical integration~(\ref{AMM3}) yields $a_{\mu}^{(1)} = 1.78 \times 10^{-9} {\rm{GeV}}.$ Using the experimental imprecision of the measurement given in~(\ref{amuhad}), we impose $\frac{2\bar{g}\xi_0}{g_{\rho\pi\pi}} a_{\mu}^{(1)} < 60 \times 10^{-11},$ which yields the following upper bound:
\begin{equation}
\bar{g}\xi_0 < 1.0 ~{\rm{GeV}}^{-1}.
\end{equation}

Note that this upper bound is due to the experimental imprecision of the measurement~(\ref{amuhad}) and the smallness of the LV contribution for the anomalous magnetic moment, $a_{\mu}^{(1)}$.  Although two-loop corrections for the $\rho$ propagator yield smaller contributions to $a_{\mu}^{(1)}$, it will not improve the first order upper bounds, since they involve second order LV coefficients, as $(\bar{g}\xi_0)^2$. It is possible to consider vertex corrections, which contribute at the same order as the one loop corrections. But, as the vertex corrections do not affect the form factor in the timelike region \cite{Dominguez:2007dm}, do not need to be taken into account. 

Finally, as this bound is stated on the zeroth component of the LV background using data belonging to the center of mass frame (LAB), it is equally written in the Sun's frame.

\section{Conclusion and final remarks}
\label{sec:CONCLUSION}

In this work we have examined the KLZ model in the presence of LV nonminimal
couplings, named here as LV KLZ. We started the section~\ref{sec:LVVMD} revising the VMD in the model by introducing the electromagnetic interactions in a gauge invariant way. We also showed the second version of VMD, valid in the universality limit $g_{\rho\pi\pi}=g_{\rho}$, where the EM current one converts entirely into a $\rho$ meson. After that, the LV is included via two nonminimal couplings: the $\bar{g}$ between the neutral vector meson field strength and the pion current and the $\tilde{g}$ between the photon field strength and the pion current. The requirement of the second version of VMD lead us to impose the new universality condition $e\bar{g}=\tilde{g}g_{\rho}$. This signalizes that any other nonminimal couplings will lead to new universality conditions due to the VMD. It is worth highlighting that this requirement makes easier our study of VMD and pion form factor, but it can be relaxed. In section~\ref{sec:LVDECAY} we used the LV KLZ model to study the $\rho$ meson decay modified by the LV nonminimal couplings. At tree level, we have found that LV brings a linear correction in the $\rho$ meson decay, like, $\xi^{\mu} q_{\mu}$. Working in the $\rho$ rest frame, which matches with the CM frame, and using the experimental values for the parameters of the model, we also have found the quantity $\approx 1.9\times10^{-2}~\left(\mathrm{{GeV}}\right)\mathrm{^{-1}}$ as upper limit for the LV nonminimal coupling $\bar{g}\xi_{0}$ .

In section~\ref{sec:LVPIONFF}, using the VMD within of the LV KLZ model, we have found that the LV nonminimal coupling brings an anisotropy for the timelike pion form factor. To deal with pole at the $\rho$ mass, we evaluated the LV correction for the vacuum polarization and rewrite the form factor just at LV first order. At low momenta, we have noticed that the LV term brings a linear momentum dependence ($\xi^{\mu}q_{\mu}$). This fact could originate a nonzero pion mean radius $\langle r_{\pi}\rangle$ and, as a consequence, a electric dipole momentum for pion caused by the LV term, which would open a new perspective on the pion. We left this study for a future work. On the other hand, we also have noticed that LV term does not cause any effect on the pion mean-squared radius $\langle r_{\pi}^{2} \rangle$, coefficient of the $q^2$, and the reason is because we just considered LV first order corrections for the pion form factor. Taking LV higher order corrections we expect small corrections for pion mean-squared radius. To constrain the LV nonminimal coupling, we have used the two-pion contribution for the muon anomalous magnetic moment~($a_{\mu}^{had,\pi}$): In the center of mass frame, where just the time component of $\xi^{\mu}$ survives, we found the LV first order correction for $a_{\mu}^{had,\pi}$ and using the experimental imprecision of the measurement for this quantity we found $\approx 1.01 \times~{\rm{GeV}}^{-1}$ as upper limit for $\bar{g}\xi_0$, which is greater than found using the imprecision of the $\rho$ decay constant.


As a final remark we pointed that the LV nonminimal coupling between vector mesons and pions, introduced in the KLZ model at hadron level, can be generated from Lorentz violating effects at quark level described by a QCD dimension-five and dimension-six Lagrangian. At this level, the LV terms would consist of extra interactions between gluons and quarks, making the theory nonrenormalizable. The effective lagrangian at hadron level with LV nonminimal coupling is constructed by matching the relevant symmetry
properties at quark level, in particular the chiral symmetry. This way, following~\cite{Quiral,Pospelov2,Noordmans2}, one can derive the chiral perturbation theory (CPTh) corrected by the LV nonminimal contributions. In addition, the vector mesons are implemented in the CPTh by a residual hidden local symmetry mechanism~\cite{Bando:1984ej}. It seems that the implementation of the LV terms does not have influence under this mechanism. Nonminimal couplings at the quark level can be also relevant to investigate nucleon contributions to hadron and nuclear EDM, where the CP violation is a sensitive issue \cite{Pospelov3}. As the KLZ model is just the CPTh with vector mesons expanded up to second order, the present work opens a window to explore the LV nonminimal couplings within of the chiral perturbation theory and at the quark level as well, which is very rich for describing the hadron physics.  Another interesting perspective is the investigation of CP-odd pion-nucleon LV interactions, both in the minimimal and nonminimal mass dimension, which can play a relevant role as source of nuclear EDM \cite{Pospelov3,Yamanaka,EDM3}.


\section{References}

\begin{thebibliography}{99}                                                                                               %


\bibitem {Weise:1974bk}W.~Weise,
Phys.\ Rept.\ \textbf{13}, 53 (1974). doi:10.1016/0370-1573(74)90005-2


\bibitem {Jegerlehner:2009ry}F.~Jegerlehner and A.~Nyffeler,
Phys.\ Rept.\ \textbf{477}, 1 (2009) doi:10.1016/j.physrep.2009.04.003
[arXiv:0902.3360 [hep-ph]].


\bibitem {Kendall:1991np}H.~W.~Kendall,
Rev.\ Mod.\ Phys.\ \textbf{63}, 597 (1991). doi:10.1103/RevModPhys.63.597


\bibitem {Leupold:2009kz}S.~Leupold, V.~Metag and U.~Mosel,
Int.\ J.\ Mod.\ Phys.\ E \textbf{19}, 147 (2010) doi:10.1142/S0218301310014728
[arXiv:0907.2388 [nucl-th]].


\bibitem {Sakurai}J.~J.~Sakurai, Currents and Mesons, University of Chicago
Press (1969).

\bibitem {Sakurai:1960ju}J.~J.~Sakurai,
Annals Phys.\ \textbf{11}, 1 (1960). doi:10.1016/0003-4916(60)90126-3

\bibitem {Kroll:1967it}N.~M.~Kroll, T.~D.~Lee and B.~Zumino,
Phys.\ Rev.\ \textbf{157}, 1376 (1967). doi:10.1103/PhysRev.157.1376


\bibitem {vanHees:2003dk}H.~van Hees,
hep-th/0305076.


\bibitem {Ruegg:2003ps}H.~Ruegg and M.~Ruiz-Altaba,
Int.\ J.\ Mod.\ Phys.\ A \textbf{19}, 3265 (2004)
doi:10.1142/S0217751X04019755 [hep-th/0304245].


\bibitem {Gale:1990pn}C.~Gale and J.~I.~Kapusta,
Nucl.\ Phys.\ B \textbf{357}, 65 (1991). doi:10.1016/0550-3213(91)90459-B

\bibitem {Gounaris:1968mw}G.~J.~Gounaris and J.~J.~Sakurai,
Phys.\ Rev.\ Lett.\ \textbf{21}, 244 (1968). doi:10.1103/PhysRevLett.21.244


\bibitem {Davier:2005xq}M.~Davier, A.~Hocker and Z.~Zhang,
Rev.\ Mod.\ Phys.\ \textbf{78}, 1043 (2006) doi:10.1103/RevModPhys.78.1043
[hep-ph/0507078].

\bibitem{deTroconiz:2004yzs}
  J.~F.~de Troconiz and F.~J.~Yndurain,
  Phys.\ Rev.\ D {\bf 71}, 073008 (2005)
  doi:10.1103/PhysRevD.71.073008
  [hep-ph/0402285].

\bibitem{Brown:2001mga}
  H.~N.~Brown {\it et al.} [Muon g-2 Collaboration],
  Phys.\ Rev.\ Lett.\  {\bf 86}, 2227 (2001)
  doi:10.1103/PhysRevLett.86.2227
  [hep-ex/0102017].

\bibitem{Kinoshita:2004wi}
  T.~Kinoshita and M.~Nio,
  Phys.\ Rev.\ D {\bf 70}, 113001 (2004)
  doi:10.1103/PhysRevD.70.113001
  [hep-ph/0402206].

\bibitem{Jackiw:1972jz}
  R.~Jackiw and S.~Weinberg,
  Phys.\ Rev.\ D {\bf 5}, 2396 (1972).
  doi:10.1103/PhysRevD.5.2396

\bibitem {Patrignani:2016xqp}C.~Patrignani \textit{et al.} [Particle Data
Group],
Chin.\ Phys.\ C \textbf{40}, no. 10, 100001 (2016).
doi:10.1088/1674-1137/40/10/100001

\bibitem {Dominguez:2007dm}C.~A.~Dominguez, J.~I.~Jottar, M.~Loewe and
B.~Willers,
Phys.\ Rev.\ D \textbf{76}, 095002 (2007) doi:10.1103/PhysRevD.76.095002
[arXiv:0705.1902 [hep-ph]].


\bibitem {Colladay}D.~Colladay and V.A.~Kosteleck\'{y}, Phys. Rev.
D~\textbf{55}, 6760 (1997); D.~Colladay and V.A.~Kosteleck\'{y}, Phys. Rev.
D~\textbf{58}, 116002 (1998).

\bibitem {Fermion1}V.A.~Kosteleck\'{y} and C.D.~Lane, J. Math.
Phys.~\textbf{40}, 6245 (1999); V.A.~Kosteleck\'{y} and R.~Lehnert, Phys. Rev.
D~\textbf{63}, 065008 (2001); D.~Colladay and V.A.~Kosteleck\'{y}, Phys. Lett.
B~\textbf{511}, 209 (2001); R.~Lehnert, Phys. Rev. D \textbf{68}, 085003
(2003); R.~Lehnert, J. Math. Phys.~\textbf{45}, 3399 (2004); B.~Altschul,
Phys. Rev. D~\textbf{70}, 056005 (2004).

\bibitem {Fermion2}W.F.~Chen and G.~Kunstatter, Phys. Rev. D~\textbf{62},
105029 (2000); O.G.~Kharlanov and V.Ch.~Zhukovsky, J. Math. Phys.~\textbf{48},
092302 (2007); B.~Gon\c{c}alves, Y.N.~Obukhov, and I.L.~Shapiro, Phys. Rev.
D~\textbf{80}, 125034 (2009); S.I. Kruglov, Phys. Lett. B~\textbf{718}, 228
(2012); T.J.~Yoder and G.S.~Adkins, Phys. Rev. D~\textbf{86}, 116005 (2012);
S.~Aghababaei, M.~Haghighat, I.~Motie, Phys. Rev. D~\textbf{96}, 115028 (2017).

\bibitem {Adam1}S.M. Carroll, G.B.~Field, and R.~Jackiw, Phys. Rev.
D~\textbf{41}, 1231 (1990);  C.~Adam and F.R.~Klinkhamer, Nucl. Phys.
B~\textbf{607}, 247 (2001);  Nucl. Phys. B~\textbf{657}, 214 (2003); V.Ch.~Zhukovsky, A.E.~Lobanov, and
E.M.~Murchikova, Phys. Rev. D~\textbf{73}, 065016 (2006); J.~Alfaro, A.A.~Andrianov, M.~Cambiaso, P.~Giacconi, and
R.~Soldati, Int. J. Mod. Phys. A~\textbf{25}, 3271 (2010); Y.M.P.~Gomes,
P.C.~Malta, Phys. Rev. D~\textbf{94}, 025031 (2016); A.~Mart\'{\i}n-Ruiz,
C.A.~Escobar, Phys. Rev. D~\textbf{95}, 036011 (2017); T. R. S. Santos and R. F. Sobreiro, Phys. Rev. D
\textbf{91}, 025008 (2015); Braz. J. Phys. \textbf{46}, 437 (2016).

\bibitem {KM}V.A.~Kosteleck\'{y} and M.~Mewes, Phys. Rev. Lett.~\textbf{87},
251304 (2001);  Phys. Rev. D~\textbf{66},056005 (2002); Phys. Rev. Lett.~\textbf{97},
140401 (2006).

\bibitem {Escobar}C.A.~Escobar and M.A.G.~Garcia, Phys. Rev. D~\textbf{92},
025034 (2015); A.~Mart\'{\i}n-Ruiz and C.A.~Escobar, Phys. Rev. D~\textbf{94},
076010 (2016).

\bibitem {KFint}B.~Altschul, Phys. Rev. Lett.~\textbf{98}, 041603 (2007);
C.~Kaufhold and F.R.~Klinkhamer, Phys. Rev. D~\textbf{76}, 025024 (2007);
F.R.~Klinkhamer and M.~Risse, Phys. Rev. D~\textbf{77}, 016002 (2008);
Phys. Rev. D~\textbf{77}, 117901 (2008);
F.R.~Klinkhamer and M.~Schreck, Phys. Rev. D~\textbf{78}, 085026 (2008).

\bibitem {Interac}A.~Moyotl, H.~Novales-S\'{a}nchez, J.J.~Toscano, and
E.S.~Tututi, Int. J. Mod. Phys. A~\textbf{29}, 1450039 (2014); Int. J. Mod.
Phys. A~\textbf{29}, 1450107 (2014).
\bibitem {Vertex} R. Bufalo, Int. J. Mod. Phys. A \textbf{29},
1450112 (2014); G. P. de Brito,  P. C. Malta, and L. P. R. Ospedal, Phys. Rev. D \textbf{95}, 016006 (2017).

\bibitem {QED} G. P. de Brito, J. T. Guaitolini Jr, D. Kroff, P. C. Malta, and C. Marques, Phys. Rev. D \textbf{94} 056005 (2016); T. R. S. Santos, R. F. Sobreiro, Phys. Rev. D \textbf{94}, 125020 (2016); A. F. Santos, and F. C. Khanna, Phys. Rev. D \textbf{95},25012 (2017); Adv. High E. Phys. 2018, 4596129 (2018); F.E.P. dos Santos, M. M. Ferreira Jr, Symmetry \textbf{10}, 302 (2018); T. P. Netto, Phys.Rev. D\textbf{97}, 055048 (2018); J. R. Nascimento, A. Yu. Petrov, Carlos M. Reyes, Eur. Phys. J. C \textbf{78}, 541 (2018).

\bibitem {Quiral}R. Kamand, B. Altschul, and M. R. Schindler, Phys. Rev.
D~\textbf{95}, 056005 (2017);  Phys. Rev.D~\textbf{97}, 095027 (2018).

\bibitem{Noordmans}J.P. Noordmans, Phys.Rev. D \textbf{95}, 075030 (2017); J.P. Noordmans, J.Phys.Conf.Ser. 873, 012009 (2017); J.Phys.Conf.Ser. 952, 012020 (2018).

\bibitem{Noordmans2} J.P. Noordmans, J. de Vries, R.G.E. Timmermans  Phys.Rev. C \textbf{94},  025502 (2016

\bibitem {Kostelec1}V.A.~Kosteleck\'{y} and M.~Mewes, Phys. Rev.
D~\textbf{80}, 015020 (2009); M.~Mewes, Phys. Rev. D~\textbf{85}, 116012
(2012); M.~Schreck, Phys. Rev. D~\textbf{89}, 105019 (2014).

\bibitem {Kostelec2}V.A.~Kosteleck\'{y} and M.~Mewes, Phys. Rev.
D~\textbf{88}, 096006 (2013); M.~Schreck, Phys. Rev. D~\textbf{90}, 085025 (2014).

\bibitem {Myers1}R.C.~Myers and M.~Pospelov, Phys. Rev. Lett.~\textbf{90},
211601 (2003); C.M.~Reyes, L.F.~Urrutia, and J.D.~Vergara, Phys. Rev.
D~\textbf{78}, 125011 (2008); J.~Lopez-Sarrion and C.M.~Reyes, Eur. Phys. J.
C~\textbf{72}, 2150 (2012).

\bibitem {Marat}C.M.~Reyes, L.F.~Urrutia, and J.D.~Vergara, Phys. Lett.
B~\textbf{675}, 336 (2009); C.M.~Reyes, Phys. Rev. D~\textbf{82}, 125036
(2010); C.M.~Reyes, Phys. Rev. D~\textbf{87}, 125028 (2013).

\bibitem {NModd1}H.~Belich, T.~Costa-Soares, M.M.~Ferreira, Jr., and
J.A.~Helay\"{e}l-Neto, Eur. Phys. J. C~\textbf{41}, 421 (2005); H.~Belich,
L.P.~Colatto, T.~Costa-Soares, J.A.~Helay\"{e}l-Neto, and M.T.D.~Orlando, Eur.
Phys. J. C~\textbf{62}, 425 (2009);

\bibitem {NMmaluf}B. Charneski, M. Gomes, R. V. Maluf, and A. J. da Silva,
Phys. Rev. D \textbf{86}, 045003 (2012); A.F.~Santos, and Faqir~C.~Khanna,
Phys. Rev. D~\textbf{95}, 125012 (2017).

\bibitem {Radio}G. Gazzola, H. G. Fargnoli, A. P. Baeta Scarpelli, M. Sampaio,
and M. C. Nemes, J. Phys. G \textbf{39}, 035002 (2012); A. P. Baeta Scarpelli,
J. Phys. G \textbf{39}, 125001 (2012).

\bibitem {NMbakke}K. Bakke, H. Belich, and E. O. Silva, J. Math. Phys.
\textbf{52}, 063505 (2011); \ J. Phys. G \textbf{39}, 055004 (2012); Annalen
der Physik (Leipzig) \textbf{523}, 910 (2011);  L. H. C. Borges, A. G. Dias, A. F. Ferrari, J. R. Nascimento, A. Yu. Petrov, Phys. Lett. B 756, 332 (2016); Y.M.P. Gomes, J.T. Guaitolini Jr, arXiv:1808.02029 .

\bibitem {Pospelov}P. A. Bolokhov, M. Pospelov, and M. Romalis, Phys. Rev. D
\textbf{78}, 057702 (2008).

\bibitem {Jonas1}J. B. Araujo, R. Casana and M.M. Ferreira, Jr., Phys. Rev. D
\textbf{92}, 025049 (2015); Phys. Lett. B \textbf{760}, 302 (2016).

\bibitem {Pospelov2}P. A. Bolokhov, M. Pospelov, and M. Romalis, Phys. Rev. D \textbf{77}, 025022 (2008).

\bibitem {Koste2}Y. Ding, V.A. Kostelecky, Phys. Rev. D  \textbf{94}, 056008 (2016).

\bibitem {Victor}V. E. Mouchrek-Santos and M.M. Ferreira, Jr., Phys. Rev. D
\textbf{95}, 071701(R) (2017).

\bibitem {Josberg}R. Casana, J. S. Rodrigues, F. E. P. dos Santos, Phys. Lett. B \textbf{790}, 354 (2019).

\bibitem {Koste3}  V.A. Kostelecky and Z. Li, "Gauge field theories with Lorentz-violating operators of arbitrary dimension", arXiv:1812.11672 .

\bibitem {Jonas4} J. B. Araujo, A. H. Blin, M. Sampaio, M. M. Ferreira Jr,
"Constraining dimension-6 nonminimal Lorentz-violating electron-nucleon interactions with EDM physics", arXiv:1902.10329.

\bibitem {Sideral}R. Bluhm, V. Alan Kostelecky, C. D. Lane, and N. Russel,
Phys. Rev. Lett. \textbf{88},\ 090801 (2002); Phys. Rev. D \textbf{68},
125008\ (2003); V.A. Kostelecky and\ M.\ Mewes,\ Phys.\ Rev.\ D\ \textbf{66},\ 056005\ (2002).

\bibitem {Lowenstein:1972pr}J.~H.~Lowenstein and B.~Schroer,
Phys.\ Rev.\ D \textbf{6}, 1553 (1972). doi:10.1103/PhysRevD.6.1553

\bibitem{Djukanovic:2004mm}
  D.~Djukanovic, M.~R.~Schindler, J.~Gegelia, G.~Japaridze and S.~Scherer,
  Phys.\ Rev.\ Lett.\  {\bf 93}, 122002 (2004)
  doi:10.1103/PhysRevLett.93.122002
  [hep-ph/0407239].

\bibitem {Pion}B. Altschul, Phys. Rev. D 84, 091902(R) (2011); 87, 096004
(2013); 88, 076015 (2013).

\bibitem{Lomon:2016eyp}
  E.~L.~Lomon and S.~Pacetti,
  Phys.\ Rev.\ D {\bf 94}, no. 5, 056002 (2016)
  doi:10.1103/PhysRevD.94.056002
  [arXiv:1603.09527 [hep-ph]].

\bibitem{Dominguez:2017gmc}
  C.~A.~Dominguez, M.~Lushozi and K.~Schilcher,
  arXiv:1710.10874 [hep-ph].

\bibitem{Bando:1984ej}
  M.~Bando, T.~Kugo, S.~Uehara, K.~Yamawaki and T.~Yanagida,
  Phys.\ Rev.\ Lett.\  {\bf 54}, 1215 (1985).
  doi:10.1103/PhysRevLett.54.1215

  \bibitem {Pospelov3}M. Pospelov and A. Ritz, Annals of Phys. \textbf{318}, 119 (2005).




  \bibitem {Yamanaka}N. Yamanaka, Int. J. Mod. Phys. E 26, 1730002 (2017); N. Yamanaka, B. Sahoo, N. Yoshinaga, T. Sato, K. Asahi, B. Das, Eur. Phys. J. A 53, 54 (2017).


 \bibitem {EDM3}J. Engel, M.J. Ramsey-Musolf and U. van Kolck, Electric dipole
moments of nucleons, nuclei and atoms: the Standard Model and beyond, Prog.
Part. Nucl. Phys. 71 (2013) 21; T. Chupp and M. Ramsey-Musolf, Electric dipole
moments: a global analysis, Phys. Rev. C 91 (2015) 035502.

\end{thebibliography}
\end{document}